\documentclass[aps,pre,twocolumn,showpacs,superscriptaddress]{revtex4}
\usepackage{amssymb}
\usepackage{bbm}
\usepackage{indentfirst}
\usepackage{graphicx}
\usepackage{amsmath,amssymb}
\usepackage{bpchem}

\usepackage{color}
\usepackage{tabularx}
\usepackage{multirow}
\usepackage{makecell}
\usepackage{array}
\usepackage{appendix}
\newcommand{\PreserveBackslash}[1]{\let\temp=\\#1\let\\=\temp}
\newcolumntype{C}[1]{>{\PreserveBackslash\centering}p{#1}}
\newcolumntype{R}[1]{>{\PreserveBackslash\raggedleft}p{#1}}
\newcolumntype{L}[1]{>{\PreserveBackslash\raggedright}p{#1}}
\begin{document}

\title{Dynamics of phase oscillators in the Kuramoto model with generalized
frequency-weighted coupling}

\author{Can Xu}
\affiliation{College of Information Science and Engineering, Huaqiao University, Xiamen 361021, China}
\affiliation{Department of Physics and the Beijing-Hong Kong-Singapore Joint Center for Nonlinear and Complex Systems (Beijing), Beijing Normal University, Beijing 100875, China}

\author{Jian Gao}
\affiliation{Department of Physics and the Beijing-Hong Kong-Singapore Joint Center for Nonlinear and Complex Systems (Beijing), Beijing Normal University, Beijing 100875, China}

\author{Hairong Xiang}
\affiliation{Department of Physics and the Beijing-Hong Kong-Singapore Joint Center for Nonlinear and Complex Systems (Beijing), Beijing Normal University, Beijing 100875, China}

\author{Wenjing Jia}
\affiliation{Department of Physics and the Beijing-Hong Kong-Singapore Joint Center for Nonlinear and Complex Systems (Beijing), Beijing Normal University, Beijing 100875, China}

\author{Shuguang Guan}
\affiliation{Department of Physics, East China Normal University, Shanghai 200241, China}

\author{Zhigang Zheng}\email{zgzheng@hqu.edu.cn}
\affiliation{College of Information Science and Engineering, Huaqiao University, Xiamen 361021, China}

\begin{abstract}
{\bf We generalize the Kuramoto model for the synchronization transition of globally coupled phase oscillators to populations by incorporating an additional heterogeneity with the coupling strength, where each oscillator  pair interacts with different coupling strength weighted by a genera; function of their natural frequency. The expression for the critical coupling can be straightforwardly extended to a generalized explicit formula analytically, and s self-consistency approach is developed to predict the stationary states in the thermodynamic limit. The landau damping effect is further revealed by means of the linear stability analysis and resonance poles theory above the critical threshold which turns to be far more generic. Furthermore, the dimensionality reduction technique of the Ott-Antonsen is implemented to capture the analytical description of relaxation dynamics of the steady states valid on a globally attracting manifold. Our theoretical analysis and numerical results are consistent with each other, which can help us understand the synchronization transition in general  networks with heterogenous couplings.}
\end{abstract}


\maketitle
\section{Introduction}\label{secone}
Emergence of spontaneous synchronization in a population of interacting elements is one important issue in the nonlinear dynamics and complex networks. Investigating the intrinsic microscopic mechanism of such phenomena provides insights for understanding the collective behaviors in a wide variety of fields, such as the flashing of fireflies, circadian rhythms, electrochemical and spin-toque oscillators, applause formation in a large audience, the power grids, and then again in other real systems~\cite{pikovsky2001,strogatz2004,arenas2008,breakspear2010}. Mathematically, the most successful model for studying synchronization problem was introduced by Kuramoto~\cite{Kuramoto1975}, which stands for the classical paradigms for synchronization and turns out to be analytical solvable . During the last decades, the Kuramoto model with its generalizations have inspired and simulated extensive studies from serval aspects, including the fundamental theory analysis and their relevance to practice~\cite{acebron2005,rodrigues2016}.

The Kuramoto model describes the evolution of an ensemble of coupled phase oscillators by means of a set of time differential equations,
\begin{equation}\label{equ:re1}
\dot\theta_{i}=\omega_{i}+\dfrac{K}{N}\sum_{j=1}^{N}\sin(\theta_{j}-\theta_{i}),\qquad i=1,2,\cdots,N,
\end{equation}
where $\theta_{i}(t)$ is instantaneous phase of the $i$th oscillator, $\omega_{i}$ is its natural frequency usually extracted from a certain probability density function $g(\omega)$, and $K>0$ is the global coupling strength. Eq.(1) describes the system towards synchrony characterized by the order parameter,
\begin{equation}\label{equ:re2}
z(t)= r(t)e^{i\Psi(t)}=\dfrac{1}{N}\sum_{j=1}^{N}e^{i\theta_{j}(t)},
\end{equation}
where $z(t)$ is a complex valued vector on the complex plane, $r(t)$ is the collective amplitude, $\Psi(t)$ is its phase. It has been shown that, with the increasing of coupling strength $K$, the system transits from the incoherent state, in which $r=0$, and the oscillators evolve almost according to their natural frequency, into the coherent state, in which $r>0$, and a number of oscillators became synchronized, sharing a common effective frequency $\Omega$. As already well know results~\cite{kuramoto1984}, Kuramoto showed that in the limit $N\rightarrow \infty$, and the natural frequency distribution function $g(\omega)$ is one peaked, symmetric with respective to its center $\hat{\omega}$, a continuous second-order phase transition occurs at the critical coupling strength,
\begin{equation}\label{equ:re3}
K_{c}=\dfrac{2}{\pi g(\hat{\omega})},
\end{equation}
and the collective amplitude $r$ satisfies the $1/2$ scale-law
\begin{equation}\label{equ:re4}
r\propto\sqrt{\frac{K-K_{c}}{K_{c}}},
\end{equation}
close to the critical point $K_{c}$. With further increasing of $K$ above $K_{c}$, other oscillators may in turn be entrained by the synchronized group, forming a macroscopic oscillating cluster, and the size of cluster become larger and larger, when $K$ is large enough, eventually, all the oscillators totally coincide with each other, as the consequence, $z$ become a vector on the unit circle and $r$ approaches to $1$ characterizing the system become complete synchrony.

Remarkably, the only heterogeneity of the classical Kuramoto model lies in their natural frequency, while the dispersion of frequencies competes with the attractive coupling $K$, in a way that a phase transition to synchronization takes place when the coupling strength is strong enough, or the frequency distribution $g(\omega)$ is sufficiently narrow. A straightforward extension of the current model is to add a new component of heterogeneity into the coupling strength, which is a natural attribute in the realistic systems~\cite{zanette2005,paissan2007,paissan2008,vlasov2014,skardal2016,vlasov2015}. For examples, when the phase oscillators are set on complex networks, the network properties strongly impact the route to synchronization, moreover, this structure heterogeneity is equivalent to the coupling heterogeneity in a mean-field form under suitable approximation~\cite{yoon2015,coutinho2013}. In particular, recent studies of the traveling state, the $\pi$-state in an ensemble of coupled phase oscillators is essentially a special case of heterogeneity coupling pattern, where the interacting strength among elements has only two kinds of possible choice, either $K$ or $-K$ with different proportion coefficient~\cite{daido1996,hong2011}. Furthermore, it has been shown that the Kuramoto model with frequency-weighted coupling can lead to non-trivial dynamical consequences, such as the discontinuous transition to collective synchronization in general networks~\cite{zhang2013,leyva2013,zhang2014,hu2014,zhou2015,xu2016}, the chimera states~\cite{wang2011,zhu2013}, and the Landau damping effects in conformist and contrarian oscillators~\cite{qiu2015}.

The aim of this paper is to extend the Kuramoto theory by incorporating an additional source of heterogeneity with the coupling strength as advanced above, we present a complete framework to investigate the generalized frequency-weighted model. First, we establish rigorous self-consistency equations for the order parameter amplitude and the synchronization frequency $\Omega$, through which all the possible steady states of the system could be predicted. In contrast to the case of homogeneous coupling, the synchronization frequency is not necessarily equal to the mean value of the natural frequency, Actually, the critical frequency $\Omega_{c}$ plays an important role in determining the critical coupling strength $K_{c}$. Second, we pay our particular attentions to the relaxation dynamics of incoherent state when $K<K_{c}$, a detailed linear stability analysis of the incoherent state is performed, it is shown that the linearized operator has no discrete eigenvalues below the critical threshold. Furthermore, as the byproduct of the formulation, the explicit expression for the critical coupling strength is also obtained, which keeps the unified form for the classical Kuramoto model Eq.(3) and is consistent with the mean-field theory. The linear stability foretells that the incoherent state is only neutral stable to perturbation, nevertheless, we report an theoretical analysis and show that the relaxation to the incoherent state is indeed exponential by means of the resonance poles theory, meanwhile, the relaxation rate can be solved in a general framework. Together with the numerical simulations that verify our theoretical
analyses, in the following, we report our main results, both theoretically and numerically.

  \section{ Mean-field theory}\label{sectwo}
  The Kuramoto model with generalized frequency-weighted coupling is described by the following dynamical equations,
  \begin{equation}\label{equ:re5}
  \dot{\theta}_{i}=\omega_{i}+\dfrac{K f(\omega_{i})}{N}\sum_{j=1}^{N}\sin(\theta_{j}-\theta_{i}),\qquad i=1,\cdots,N,
  \end{equation}
  where $f(\omega_{i})$ is a real value function of the natural frequency $\omega_{i}$. The most important characteristic of the current model is that we extend the frequency dependence to a generalized function, in contrast to the previous studies of frequency-weight coupling, where $f(\omega_{i})=\omega_{i}^{\beta}$ or $|\omega_{i}|$ ~\cite{hu2014,zhou2015,xu2016,wang2011,zhu2013,qiu2015}. Eq.~(\ref{equ:re5}) defines a heterogeneous interaction pattern underlying the system, apart from the global coupling $K>0$, the coupling strength between oscillators is weighted by their frequency $\omega_{i}$, which is a reasonable consideration in the role of realistic systems. The definition Eq.~(\ref{equ:re2}) allows us to rewrite Eq.~(\ref{equ:re5}) in the mean-field form, which yields
  \begin{equation}\label{equ:re6}
  \dot{\theta}_{i}=\omega_{i}+Krf(\omega_{i})\sin(\Psi-\theta_{i}).
  \end{equation}
  Here $Krf(\omega_{i})$ can be interpreted as an effective coupling strength, Eq.~(\ref{equ:re6}) reflects that the interaction to oscillator $i$ is equivalent to a phase $\Psi$, and weighted by the effective coupling $Krf(\omega_{i})$. Note that, the effective coupling here can be either positive or negative according to the sign of weighted-function $f(\omega_{i})$. As a matter of fact, the oscillators in the ensemble can be in general grouped into two populations, when $f(\omega_{i})>0$, the coupling to the mean-field is attractive, and synchronization is triggered by these oscillators, whereas $f(\omega_{i})<0$, the interaction to the mean-field turns out to be repulsive, as a result, synchronization is suppressed.

  Since we are interested in the steady state of the system, and thus the self-consistence method turns out to be effective. In the long time limit, we assume Eq.~(\ref{equ:re5}) approaches to a stationary state, where the collective amplitude $r$ is independent of time and the mean-field phase $\Psi$ rotates uniformly with a frequency, i.e, $\Psi(t)=\Omega t+\Psi_{0}$, after an appropriate phase shift, we can set $\Psi_{0}=0$. By introducing the phase difference:
  \begin{equation}\label{equ:re7}
  \varphi_{i}=\theta_{i}-\Psi,
  \end{equation}
  Eq.~(\ref{equ:re6}) can be transformed into
  \begin{equation}\label{equ:re8}
  \dot{\varphi}_{i}=\omega_{i}-\Omega-K r f(\omega_{i})\sin(\varphi_{i}),
  \end{equation}
  in the rotating frame, the evolution of each oscillator can be thought of as resulting from the interaction with the mean-field $z$. In the thermodynamic limit $N\rightarrow\infty$, a density function $\rho(\varphi,\omega,t)$ on a $(\varphi, t)$ space, with dependence on the parameter $\omega$ is needed, where $\rho(\varphi,t,\omega)d\omega$ gives the fraction of oscillators of natural frequency $\omega$, which lie between the phase deviation $\varphi$ and $\varphi+d\varphi$ at time $t$ with the normalization condition:
  \begin{equation}\label{equ:re9}
  \int_{0}^{2\pi}\rho(\varphi, \omega, t)\,d\varphi=1,
  \end{equation}
  and $2\pi$-period in $\varphi$. Consequently, Eq.~(\ref{equ:re8}) is equivalent to the following continuity equation
  \begin{equation}\label{equ:re10}
  \dfrac{\partial\rho}{\partial t}+\dfrac{\partial}{\partial \varphi}\left(\rho\cdot(\omega-\Omega-K\cdot r f(\omega)\sin\varphi)\right)=0.
  \end{equation}
  The stationary solution of Eq.~(\ref{equ:re10}) should be discussed in two distinct cases, respectively, 
  \begin{equation}\label{equ:re11}
  \small\rho(\varphi, \omega, t)=\begin{cases}
  \delta(\varphi-\arcsin(\dfrac{\omega-\Omega}{K r f(\omega)})),\quad |\omega-\Omega|\leq K r f(\omega),\\
  \\
  \dfrac{\sqrt{(\omega-\Omega)^{2}-(K r f(\omega))^{2}}}{2\pi|\omega-\Omega-K r f(\omega)\sin\varphi|},\quad otherwise,
  \end{cases}
  \end{equation}
  the first term in Eq.~(\ref{equ:re11}) corresponds to the phase-lock oscillators, and is therefore a fixed point of Eq.~(\ref{equ:re8}), which stands for the time independent asymptotic phase deviation of a entrained phase oscillator of natural frequency $\omega$ and with respect to the synchronization $\Omega$. A further linear stability analysis of Eq.~(\ref{equ:re8}) shows that a stable fixed point satisfies
  \begin{equation}\label{equ:re12}
  \cos\varphi=\operatorname{sign}(f(\omega))\sqrt{1-\left(\dfrac{\omega-\Omega}{K r f(\omega)}\right)^{2}},
  \end{equation}
  where $\operatorname{sign}(x)$ is the sign function, i.e, $\operatorname{sign}(x)=1$ when $x\geq 0$, or $\operatorname{sign}(x)=-1$, when $x<0$. The second term of Eq.~(\ref{equ:re11}) represents the drifting oscillators, where these oscillators could not be entrained by the mean-field. Eq.~(\ref{equ:re11}) relates the stationary density function with the natural frequency $\omega$, as a result, the order parameter defined in Eq.~(\ref{equ:re2}) takes the integration form:
  \begin{equation}\label{equ:re13}
  r=\int_{-\infty}^{\infty}\int_{0}^{2\pi} g(\omega)\rho(\varphi,\omega)e^{i\varphi}d\varphi\,d\omega.
  \end{equation}
  It is easy to obtain that for the drifting oscillators, 
  \begin{equation}\label{equ:re14}
  \langle\cos\varphi\rangle=\int_{0}^{2\pi} \cos\varphi\,\rho(\varphi,\omega)d\varphi=0,
  \end{equation}
  and
  \begin{equation}\label{equ:re15}
  \begin{split}
  \langle\sin\varphi\rangle=&\int_{0}^{2\pi} \sin\varphi\,\rho(\varphi,\omega)d\varphi\\
  =&\dfrac{\omega-\Omega}{K r f(\omega)}\left[1-\sqrt{1-\left(\dfrac{K r f(\omega)}{\omega-\Omega}\right)^{2}}\right].
  \end{split}
  \end{equation}
  Taking into account the contribution of both the phase-locked and the drifting oscillators to the order-parameter, Replacing Eq.~(\ref{equ:re11}) of $\rho(\varphi,\omega)$ in Eq.~(\ref{equ:re13}), and separating real and imaginary part, yields the self-consistency equations, 
  \begin{equation}\label{equ:re16}
  \begin{split}
  r=\int_{-\infty}^{\infty}&d\omega\,g(\omega)\operatorname{sign}(f(\omega))\\
  &\sqrt{1-(\dfrac{\omega-\Omega}{K r f(\omega)})^{2}}\,\Theta(1-|\dfrac{\omega-\Omega}{K r f(\omega)}|),
  \end{split}
  \end{equation}
  for the collective amplitude $r$, and
  \begin{equation}\label{equ:re17}
  \begin{split}
  0=\int_{-\infty}^{\infty}&d\omega\,g(\omega)\dfrac{\omega-\Omega}{K r f(\omega)}-\int_{-\infty}^{\infty}d\omega\,g(\omega)\dfrac{\omega-\Omega}{K r f(\omega)}\\
  &\sqrt{1-(\dfrac{K r f(\omega)}{\omega-\Omega})^{2}}\,\Theta(|\dfrac{\omega-\Omega}{K r f(\omega)}|-1),
  \end{split}
  \end{equation}
  for the mean-field frequency $\Omega$, here $\Theta(x)$ is the Heaviside function.

  Eq.~(\ref{equ:re16}) and Eq.~(\ref{equ:re17}) together provide a closed equation for the dependence of magnitude $r$ and the frequency $\Omega$ of the mean-field on $K$, in terms of the distribution $g(\omega)$ and weighted-function $f(\omega)$. Theoretically, an explicit expression for $r(K)$ and $\Omega(K)$ could be solved analytically or numerically with a given $g(\omega)$ and $f(\omega)$, however, a full analysis of the solution for an arbitrary form of $g(\omega)$, taking into account any possible dependence on the weighted-function $f(\omega)$ is out of reach. Correspondingly, we focus on our attentions by pointing out the representative and generic properties of the self-consistently equations, we look for the critical point with the onset of non-vanishing mean-field, hence, in the limit case $r\rightarrow 0$, taking into account the Taylor expansion of Eq.~(\ref{equ:re16}) and Eq.~(\ref{equ:re17}), thus the critical coupling strength $K_{c}$ reads
  \begin{equation}\label{equ:re18}
  K_{c}=\operatorname{sign}[f(\Omega_{c})]\dfrac{2}{\pi g(\Omega_{c})f(\Omega_{c})},
  \end{equation}
  and the critical mean-field frequency $\Omega_{c}$ satisfies the balance equation
  \begin{equation}\label{equ:re19}
  P\cdot\int_{-\infty}^{\infty}\dfrac{g(\omega)f(\omega)}{\omega-\Omega_{c}}d\omega=0,
  \end{equation}
  the symbol $P$ means the principal-value integration within the whole real line.

  The concise expression Eq.~(\ref{equ:re18}) is significant, which can be interpreted as a straightforward extension of the classical Kuramoto model Eq.~(\ref{equ:re3}), together with a critical frequency $\Omega_{c}$ determined by the balance equation Eq.~(\ref{equ:re19}). In contrast to the previous studies of Kuramoto model with heterogeneous coupling, where the weighted coupling is uncorrelated with the natural frequency, and therefore the second integration in Eq.~(\ref{equ:re17}) vanishes, consequently, the mean-field frequency $\Omega$ is always equal to the center $\hat{\omega}$, provided that $g(\omega)$ is one-humped and $g(\hat{\omega}-x)=g(\hat{\omega}+x)$. For the current model, however, there is no guarantee that $\Omega$ is necessary equal to $\hat{\omega}$ due to the rotational symmetry-breaking of the system~\cite{basnarkov2008,petkoski2013}. Actually, $\Omega_{c}$ plays a crucial role in determining the coupling strength, since $K_{c}$ is inverse proportional to the value $g(\Omega_{c})$ and $f(\Omega_{c})$.

\begin{table*}[htp!]\footnotesize\label{Tab:01}
\renewcommand\arraystretch{2.3}

\begin{center}
\caption{Summary of the weighted function $f(\omega_{i})$, the frequency distributions $g(\omega)$, the balance equations, the critical mean-field frequencies $\Omega_{c}$, and the critical coupling strength $K_{c}$.}
\begin{tabular}{C{2cm}C{4cm}R{5cm}R{3cm}C{3cm}}
  \hline
  \hline
  \makecell[l]{\textbf{Weight Function}} & \makecell[c]{\textbf{FD}} & \makecell[r]{\textbf{Balance Equation}} & \makecell[r]{$\bf{\Omega_c}$} & \makecell[c]{$\bf{K_c}$} \\
  \hline

  $|\omega_{i}|$ & $\dfrac{\gamma}{\pi}\dfrac{1}{\omega^{2}+\gamma^{2}}$ & $\dfrac{2\gamma}{\pi}\Omega_{c} \ln\dfrac{\gamma}{\Omega_{c}}/(\gamma^{2}+\Omega_{c}^{2})=0$ & $0,\,\pm\gamma$ & $4$ \\\hline
  $|\omega_{i}|$ & $\dfrac{1}{2a}\Theta(a-|\omega|)$ & $\dfrac{\Omega_{c}}{2a}\ln\dfrac{a^{2}-\Omega_{c}^{2}}{\Omega_{c}^{2}}=0$ & $0,\,\pm\dfrac{a}{\sqrt{2}}$ & $\dfrac{4\sqrt{2}}{\pi}$ \\\hline
  $\omega_{i}$ & $\dfrac{\gamma}{\pi}\dfrac{1}{(\omega-\Delta)^{2}+\gamma^{2}}$ & $\dfrac{\gamma^{2}+\Delta(\Delta-\Omega_{c})}{\gamma^{2}+(\Delta-\Omega_{c})^{2}}=0$ & $\dfrac{\gamma^{2}+\Delta^{2}}{\Delta}$ & $\operatorname{sign}(\Delta)\dfrac{2\gamma}{\Delta}$ \\\hline
  $\omega_{i}$ & $\dfrac{1}{2}\Theta(1-|\omega|)$ & $1-\Omega_{c}\operatorname{arctanh}(\Omega_{c})=0$ & $\pm0.8335$ & 1.528 \\\hline
  $\omega_{i}^{2}$ & $\dfrac{\sqrt{2}\gamma^{3}}{\pi}\dfrac{1}{\omega^{4}+\gamma^{4}}$ & $\gamma^{2}(\gamma-\Omega_{c})\Omega_{c}(\gamma+\Omega_{c})=0$ & $0,\,\pm\gamma$ & $\dfrac{2\sqrt{2}}{\gamma}$ \\\hline
  $\omega_{i}^{2}$ & $\dfrac{1}{2}\Theta(1-|\omega|)$ & $\Omega_{c}-\Omega_{c}^{2}\ln(\dfrac{1+\Omega_{c}}{1-\Omega_{c}})/2=0$ & $0$,$\pm 0.8335$ & 1.8327 \\\hline
  $\dfrac{1}{\omega_{i}}$ & $\dfrac{1}{2a}\Theta(\omega-a)\Theta(2a-\omega)$ & $-\ln(-2-\dfrac{2a}{-2a+\Omega_{c}})/2a\Omega_{c}=0$ & $\dfrac{4a}{3}$ & $\dfrac{16a^{2}}{3\pi}$ \\\hline
  $\dfrac{1}{1+|\omega_{i}|}$ & $\dfrac{\gamma}{\pi}\dfrac{1}{\omega^{2}+\gamma^{2}}$ & $-\dfrac{\gamma(\gamma^{2}+\Omega_{c})}{(\gamma+\gamma^{3})(\gamma^{2}+\Omega_{c}^{2})}=0$ & $-\gamma^{2}$ & $2\gamma(1+\gamma^{2})^{2}$ \\\hline
  \multirow{2}{*}{$\dfrac{1}{1+\omega_{i}^{2}}$} & \multirow{2}{*}{$\dfrac{1}{\pi}\dfrac{1}{(\omega+\Delta)^{2}+1}$} & \multirow{2}{*}{$-\dfrac{(\Delta-2\gamma)(-3+(\Delta-\Omega_{c})\cdot\Omega_{c})}{(4+\Delta^{2})(1+(\Delta-\Omega_{c})^{2})(1+\Omega_{c}^{2})}=0$} & $\Delta/2$ & $(4+\Delta^{2})^{2}/8$ \\
  & & & $(\Delta\pm\sqrt{-12+\Delta^{2}})/2$ & $2(4+\Delta^{2}),\, \Delta^{2}\geq 12$ \\
  \hline
  \hline
\end{tabular}
\end{center}
\end{table*}

  \section{ Relaxation dynamic of the incoherent state.}\label{secthree}
{\em \textbf{Linear stability analysis}}, the analysis above reveals that all the equilibrium states for the system, as well as the collective dynamical properties of the synchronization can be characterized in the frame of mean-field theory, the collective amplitude $r$ and the group velocity $\Omega$ can be formally solved for general weighted-function through the self-consistency equations. However, a through stability of all the possible steady states are still elusive, in the following, we pay our particular attention to the incoherent state, $r\equiv 0$, and conduct a detailed linear stability analysis of it, as it will appear momentarily, the stability analysis can make up for the limitation of the mean-field theory~\cite{strogatz1991}.

Let us turn to the original phase reference, and introduce the phase density function $F(\theta, \omega, t)\equiv \rho(\varphi+\Omega t, \omega, t)$ which satisfies the continuity equation:
\begin{equation}\label{equ:re20}
\dfrac{\partial F}{\partial t}+\dfrac{\partial}{\partial \theta}(F\cdot v)=0,
\end{equation}
the velocity is given by
\begin{equation}\label{equ:re21}
v=\omega+\dfrac{K}{2 i}f(\omega)(z(t)e^{-i\theta}-z^{*}(t)e^{i\theta}),
\end{equation}
here "$*$"denotes the complex conjugate of $z(t)$, Taking into account the $2\pi$-period of $\theta$ in $F(\theta, \omega, t)$, let
\begin{equation}\label{equ:re22}
z_{n}(\omega, t)=\int_{0}^{2\pi}e^{ni\theta}F(\theta, t, \omega)d\theta,\quad n=0,1,2,\cdots,
\end{equation}
be the $n$-th Fourier coefficient of $F(\theta, t, \omega)$, following this definition, we have $z_{0}(t, \omega)\equiv 1$, and the evolution of $z_{n}(t,\omega)$ satisfies the differential equations,
\begin{equation}\label{equ:re23}
\dfrac{d z_{n}}{dt}=ni\omega z_{n}+\dfrac{nK}{2}f(\omega)(z(t)z_{n-1}-z^{*}(t)z_{n+1}),
\end{equation}
which must be solved self-consistently with an equation,
\begin{equation}\label{equ:re24}
z(t)=\int_{-\infty}^{\infty}z_{1}(t, \omega)g(\omega)d\omega.
\end{equation}
A careful examination of Eq.~(\ref{equ:re23}) reveals that, $z_{n}(t, \omega)\equiv 0$, i.e, the incoherent state, where $F(\theta, t, \omega)=1/2\pi$ and $r\equiv 0$, is always a trivial fixed point of Eq.~(\ref{equ:re23}). To study the stability of this steady state, we can consider the evolution of a weak perturbation away from the incoherent state, excluding second and higher-order terms of $\delta z_{n}$, we obtain a set of linear equations for $\delta z_{n}$, which yield, 
\begin{equation}\label{equ:re25}
\begin{cases}
\dfrac{d\delta z_{1}}{dt}=(i\omega+\dfrac{K}{2}f(\omega)\hat{P})\delta z_{1}=\hat{T}\cdot \delta z_{1}\\
\\
\dfrac{d\delta z_{n}}{dt}=ni\omega\delta z_{n},\qquad n>1.
\end{cases}
\end{equation}
here $\hat{P}$ is a linear operator defined as
\begin{equation}\label{equ:re26}
\begin{split}
\hat{P}q(\omega)=&\int_{-\infty}^{\infty}q(\omega)g(\omega)d\omega\\
=&(q(\omega), P_{0}),
\end{split}
\end{equation}
where $q(\omega)$ is a function in the weighted-Lebesgue space, $P_{0}\equiv 1$, and $( \,,\, )$ is the concise inner product notation for the integration over $\omega$ with respect to the weighted factor $g(\omega)$. Since the higher Fourier harmonics have no contribution to the order parameter $z(t)$, we are just concerned with the evolution of $\delta z_{1}(t, \omega)$, let $\lambda$ to be the eigenvalue of linear operator $\hat{T}$, we have
\begin{equation}\label{equ:re27}
\dfrac{d \delta z_{1}}{dt}=\hat{T}\cdot \delta z_{1}=\lambda\delta z_{1}.
\end{equation}
Substituting the expression of $\hat{T}$ into Eq.~(\ref{equ:re27}), multiplying both side by the inverse operator $(\lambda-i\omega)^{-1}$, we obtain
\begin{equation}\label{equ:re28}
\delta z_{1}=(\lambda-i\omega)^{-1}\dfrac{K}{2}f(\omega)(\delta z_{1}. P_{0}),
\end{equation}
Taking the inner product with $P_{0}$ for both sides, then the self-consistent eigenvalue equation for the linear operator $\hat{T}$ takes the form: 
\begin{equation}\label{equ:re29}
\int_{-\infty}^{\infty}\dfrac{f(\omega)}{\lambda-i\omega}g(\omega)d\omega=\dfrac{2}{K}, \quad \lambda \in C \setminus i\omega,
\end{equation}
where $\lambda$ is on the complex plane except for those points $i\omega$, Notice that Eq.~(\ref{equ:re29}) relates implicitly the global coupling strength $K$ with the eigenvalue $\lambda$, to simplify the discussion, we rewrite Eq.~(\ref{equ:re29}) into two equations by setting $\lambda=x+iy$, i.e,
\begin{equation}\label{equ:re30}
\int_{-\infty}^{\infty}\dfrac{x}{x^{2}+(\omega-y)^{2}}f(\omega)g(\omega)d\omega=\dfrac{2}{K},
\end{equation}
and
\begin{equation}\label{equ:re31}
\int_{-\infty}^{\infty}\dfrac{\omega-y}{x^{2}+(\omega-y)^{2}}f(\omega)g(\omega)d\omega=0.
\end{equation}
The sign of $x$ determines the stability of the incoherent state, furthermore, it has been proven that~\cite{chiba2011,chiba2015}, if the global coupling strength $K>0$ but is sufficiently small, the eigenvalue $\lambda$ of linear operator $\hat{T}$ actually dose not exist, provided that the product $f(\omega)g(\omega)$ is analytic and has no singularity on the real axis $\omega$. Accordingly, the incoherent state is only neutrally stable to perturbation in the regime $K<K_{c}$, where the operator $\hat{T}$ has only continuous spectrum $i\omega$ on the whole imaginary axis, however, as $K$ further increases, the discrete eigenvalues emerge with real part $x\neq 0$ once $K>K_{c}$. Imposing the critical condition $x\rightarrow 0^{\pm}$, $y\rightarrow y_{j}$ for Eq.~(\ref{equ:re30}), once again we obtain the critical coupling strength as
\begin{equation}\label{equ:re32}
K_{c}=\operatorname{sign}[f(y_{j})]\dfrac{2}{\pi \sup_{j}g(y_{j})f(y_{j})},
\end{equation}
where $y_{j}$ are determined by the Eq.~(\ref{equ:re31}) with the limit $x\rightarrow 0^{\pm}$. Just as mentioned in~\cite{petkoski2013}, $\Omega_{c}$ is indeed the imaginary part of the eigenvalue of operator $\hat{T}$ at the boundary of stability. Since Eq.~(\ref{equ:re31}) may hare more than one root in the limit $x\rightarrow 0^{\pm}$, $\sup_{j}$ means that we choose the $j$-th root $y_{j}$, so that the product $g(y_{j})f(y_{j})$ is maximal. Table I summarizes the balance equation, the critical mean-field frequency $\Omega_{c}$, and the critical coupling strength $K_{c}$ with respect to different frequency distributions $g(\omega)$ and weighted-function $f(\omega)$. These analytical results were supported by the previous numerical simulations~\cite{hu2014,xu2016,qiu2015}.

It should be pointed out that, in contrast to the previous discussions, where $f(\omega)>0$ for any $\omega$, thus, it can be confirmed from Eq.~(\ref{equ:re30}) that $x>0$ once $K>K_{c}$, which means that the incoherent state is always linear unstable as long as the global coupling strength is sufficiently large. However, for the current model, when we release the restriction about $f(\omega)$, the real part of the eigenvalue $\lambda$ may be negative even $K>K_{c}$, which implies that the incoherent state can be linear stable at some $K>K_{c}$, once the integration $f(\omega)g(\omega)/(x^{2}+(\omega-y)^{2})<0$, or equivalently, the repulsive terms dominate the coupling. Going further, the stability analysis also provides insight for the mean-field theory, in fact, when the repulsive terms prevail over the attractive terms underlying the system, the integration in Eq.~(\ref{equ:re16}) may be negative, which violates the precondition $r>0$. As a result, in this situation, the key assumption of the mean-field theory is unreasonable, the system can never approach a steady state $r>0$ that is described in the framework of mean-field theory, accordingly, synchronization is inhibited~\cite{hong2011,qiu2015}.\\

{\em \textbf{Landau damping effects}}, according to the above linear stability analysis, the incoherent state of model Eq.~(\ref{equ:re5}) is only neutral stable when $K<K_{c}$, however, as an already well-known result, \cite{strogatz1991} showed that the order parameter $r(t)$ in this regime actually decays to zero in the long time limit ($t\rightarrow \infty$) with an exponential form in the classical Kuramoto model. In this subsection, we show that such an effect is far more generic, as soon as phase oscillators coupled to the mean-field weighted by their natural frequency. Addressing this question is non-trivial since it has already been shown that the relaxation to the equilibrium is related to susceptibility of the system to external stimulations in statistical physics~\cite{coutinho2013,strogatz1992}, in particular, the decaying mechanism is remarkably similar to the famous Landau damping in plasma physics~\cite{strogatz1992,malmberg1964}. To this end, we report a general framework to determine the decay exponent, together with extensive numerical simulations that support the theoretical predictions.

\begin{figure}[first!]
  \centering
  \includegraphics[width=1.0\linewidth,height=0.88\linewidth]{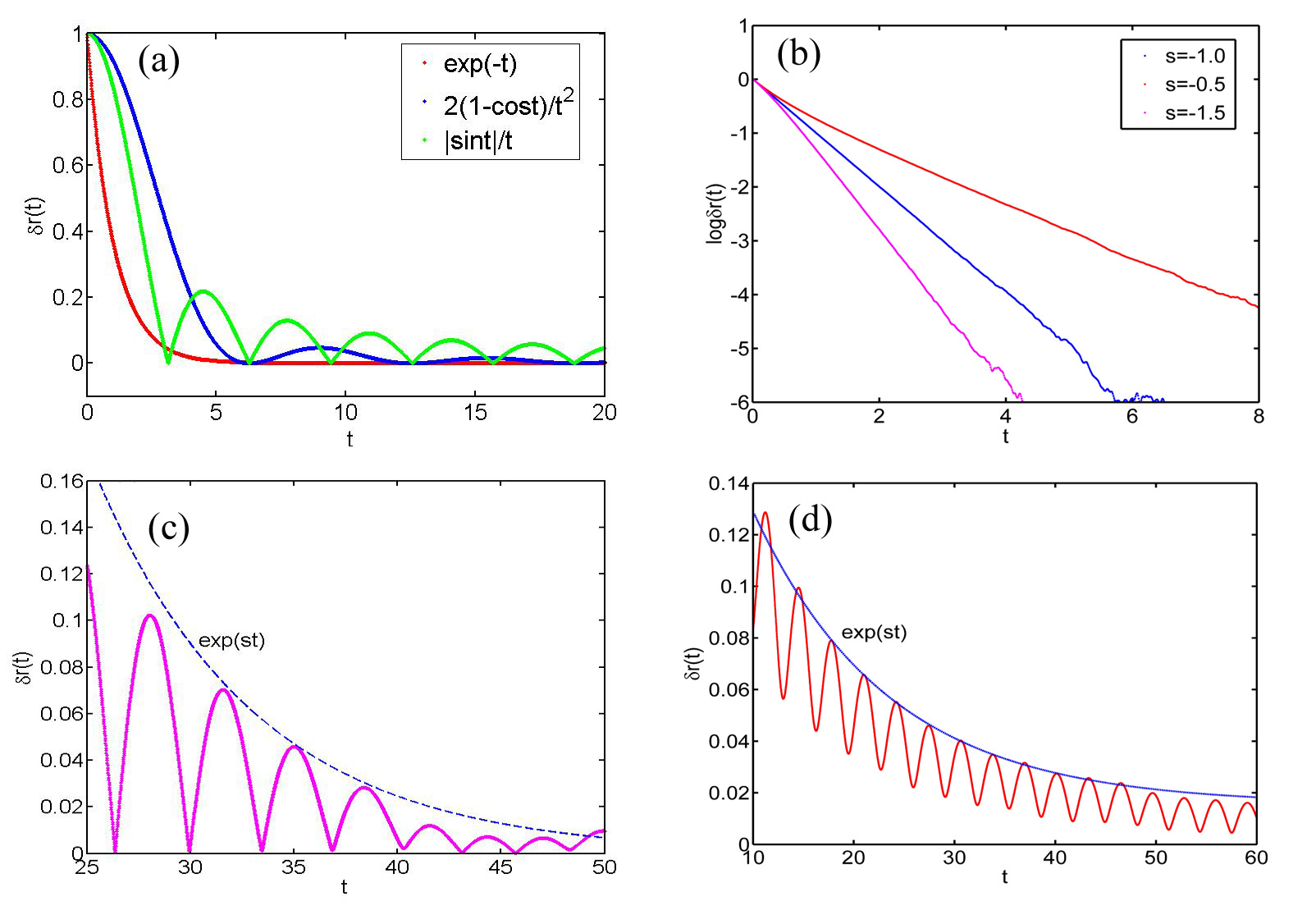}\\
  \caption{\label{first} (Color online) Different scenarios of the decay of $\delta r(t)$ with respect to different frequency distributions and weighted-functions below the critical threshold ($K<K_{c}$). Horizontal lines are time, solid lines refer to the direct numerical solutions of Eq.~(\ref{equ:re5}), and dashed lines are the fitted curve lines with exponent $s$: (a) $K=0$, the red is $g(\omega)=1/\pi(\omega^{2}+1)$, the green is $g(\omega)=1/2,\omega\in(-1,1)$ and the blue is $g(\omega)=1-|\omega|,\omega\in(-1,1)$, (b) $f(\omega)=\omega$, $g(\omega)=1/\pi((\omega-\Delta)^{2}+1)$, $K=1$ with the red $\Delta=1.0$, the blue $\Delta=0$ and the pink $\Delta=-1.0$, respectively, (c) $f(\omega)=|\omega|$,$g(\omega)=1/\pi(\omega^{2}+1)$ with $K=1.2$ and $s=-0.1298$, (d)$f(\omega)=\omega$. $g(\omega)=1/2,\omega\in(-1,1)$ with $K=0.8$ and $s=-0.07433$. All curves belong to
the neutrally stable regime of the incoherent state predicted by linear theory. In the numerical simulations, the
initial states of the system are set in the fully coherent states, and the total number of oscillators is $N=100000$}
\end{figure}

According to Eq.~(\ref{equ:re27}), a solution of $\delta z_{1}(t, \omega)$ with an initial value $\delta z_{1}(0, \omega)$ is given by
\begin{equation}\label{equ:re33}
\delta z_{1}(t, \omega)=e^{\hat{T}\cdot t}\cdot\delta z_{1}(0, \omega),
\end{equation}
where the operator $e^{\hat{T}\cdot t}$ is calculated by means of the Laplace inversion formula, which yields
\begin{equation}\label{equ:re34}
e^{\hat{T}\cdot t}=\lim_{y\rightarrow \infty}\dfrac{1}{2\pi i}\int_{x-iy}^{x+iy}e^{s\cdot t}(s-\hat{T})^{-1}ds,
\end{equation}
for $t>0$, and $x>0$. The resolvent $(s-\hat{T})^{-1}$ is obtained as follows, let $\phi(\omega)$ to be an any function in the weighted-Lebesgue space and note
\begin{equation}\label{equ:re35}
\begin{split}
\hat{R}(s)\phi=&(s-\hat{T})^{-1}\phi\\
=&(s-i\omega-\dfrac{K}{2}f(\omega)\hat{P})^{-1}\phi,
\end{split}
\end{equation}
multiplying $(s-i\omega-\dfrac{K}{2}f(\omega)\hat{P})$ for both sides, we have
\begin{equation}\label{equ:re36}
\begin{split}
(s-i\omega)\hat{R}(s)\phi=&\phi+\dfrac{K}{2}f(\omega)\hat{P}\hat{R}(s)\phi\\
=&\phi+\dfrac{K}{2}f(\omega)(\hat{R}(s)\phi,P_{0}).
\end{split}
\end{equation}
This is rearranged as
\begin{equation}\label{equ:re37}
\hat{R}(s)\phi=(s-i\omega)^{-1}\phi+\dfrac{K}{2}(\hat{R}(s)\phi,P_{0})(s-i\omega)^{-1}f(\omega),
\end{equation}
taking the inner product with $P_{0}$, we obtain
\begin{equation}\label{equ:re38}
\begin{split}
(\hat{R}(s)\phi,P_{0})=&((s-i\omega)^{-1}\phi,P_{0})+\\
&\dfrac{K}{2}(\hat{R}(s)\phi,P_{0})((s-i\omega)^{-1}f(\omega),P_{0}). 
\end{split}
\end{equation}
Reordering of the term then yields
\begin{equation}\label{equ:re39}
(\hat{R}(s)\phi,P_{0})=\dfrac{((s-i\omega)^{-1}\phi,P_{0})}{1-\frac{K}{2}((s-i\omega)^{-1}f(\omega),P_{0})}.
\end{equation}
Substituting Eq.~(\ref{equ:re39}) into Eq.~(\ref{equ:re36}) leads to
\begin{equation}\label{equ:re40}
\hat{R}(s)\phi=(s-i\omega)^{-1}\phi+\dfrac{\frac{K}{2}((s-i\omega)^{-1}\phi,P_{0})(s-i\omega)^{-1}f(\omega)}{1-\frac{K}{2}((s-i\omega)^{-1}f(\omega),P_{0})}.
\end{equation}
Then, the order parameter $z(t)$ reads
\begin{equation}\label{equ:re41}
\begin{split}
\delta z(t)=&(\delta z_{1}(t, \omega), P_{0})=(e^{\hat{T}\cdot t}\delta z_{1}(0, \omega), P_{0})\\
=&\lim_{y\rightarrow \infty}\int_{x-iy}^{x+iy}e^{st}\dfrac{D(s)}{1-K D'(s)/2}ds,
\end{split}
\end{equation}
where
\begin{equation}\label{equ:re42}
D(s)=\int_{-\infty}^{\infty}\dfrac{g(\omega)\delta z_{1}(0, \omega)}{s-i\omega}d\omega,
\end{equation}
and
\begin{equation}\label{equ:re43}
D'(s)=\int_{-\infty}^{\infty}\dfrac{g(\omega)f(\omega)}{s-i\omega}d\omega. 
\end{equation}
To invert the Laplace transform Eq.~(\ref{equ:re41}), we have to find the poles of $D(s)/(1-KD'(s)/2)$, and the calculation should be analytically continued to the left half-complex plane where $\operatorname{Re}(s)<0$. Fig.~\ref{first} presents a series of results from the numerical resolution of Eq.~(\ref{equ:re5}), in fact, from Eq.~(\ref{equ:re41}) when $K\rightarrow0$, $\delta z(t)=\int_{-\infty}^{\infty}e^{i\omega t}g(\omega)d\omega$, which is the Fourier transform of $g(\omega)$, that is, the oscillators rotate at angular frequencies given by their own natural frequencies. By the Riemann-Lebesgue lemma, we obtain $\delta z(t)\rightarrow0$ in the limit $t\rightarrow\infty$ Fig.~\ref{first}(a). When the function $g(\omega) f(\omega)$ is a rational case, such as $g(\omega)$ is Lorentzian and $f(\omega)=\omega$, the resonance poles could be solved analytically Fig.~\ref{first}(b). Also we have conducted direct numerical simulations for other typical cases, where the resonance poles could be calculated numerically, as shown in Fig.~\ref{first}(c) and (d), it is found that the envelope of the order parameter follows the form of exponential decay. The above results suggest that the Landau damping effect is a generic phenomenon which is entirely due to the occurrence of resonance poles caused by analytic continuation, and the real parts of them control the exponential relaxation rate of the the order parameter $\delta z(t)$~\cite{strogatz1991,chiba2015,strogatz1992}.

Recently, the Ott-Antonsen method has been proposed to obtain the low-dimensional dynamics of a large system of coupled oscillators~\cite{ott2008,ott2009}, the original set of differential equations can be reduced to the differential equations describing the temporal evolution of the order parameter $z(t)$ alone, which makes it possible to depict the system in a global picture. However, the validity of this method requires that, first, the $1$-th Fourier coefficient of the density function $F(\theta,\omega, t) $ can be analytically continued from the real $\omega$ into the complex $\omega$-plane, that the continuation has no singularities, second, to avoid divergence of the density function, the evolution of $z_{1}(t,\omega)$ must satisfy $|z_{1}(t,\omega)|\leq1$ at any time in the invariant manifold. In Appendix, we include the Ott-Antonsen method to study the relaxation dynamics of both the incoherent state and the coherent state~\cite{yoon2015}, provided that the restrictions above can be satisfied, it turns out that such a method is consistent with our theory formulations. For a more general case, where the preconditions are violated, the Ott-Antonsen method fail to treat properly the system, and hence it does not provide information more substantial than the traditional analysis here.  A further investigation to the system could refer to the amplitude equation theory, which reveals the local bifurcation behaviors close to the critical point when the system satisfies $O(2)$ symmetry~\cite{crawford1994,crawford1999}.

\section{Conclusion}\label{secfour}
To summarize, we extend the Kuramoto model for the synchronization transition of an infinitely large ensembles of globally coupled phase oscillators to the heterogeneously interacting pattern, where the coupled to mean-field is weighted by their natural frequency characterized by a general function $f(\omega_{i})$. Theoretically, the mean-field analysis, linear stability stability analysis, the resonance poles method, and the Ott-Antonsen reduction have been carried to obtain insights. Together with the numerical simulations, our study presented the following main results. First, we established the self-consistency equations that predicted the steady states of the system. Second, the explicit expression of the critical coupling strength was derived where the critical frequency $\Omega_{c}$ plays a crucial role in determining $K_{c}$, and it must be solved by a phase  balance equation. Third, the relaxation dynamics of the incoherent state have been addressed, and we provided the evidence of a regime ($K<K_{c}$) where the linear theory predicts neural stability, the order parameter decays exponentially, a phenomenon resembling Landau damping in plasma physics. Furthermore, the relaxation rate can be determined by the framework of resonance poles theory. Finally, we provided the Ott-Antonsen method to capture  the relaxation dynamics of a general steady state $r>0$, when the system satisfies the scope of applications. All the results are consistent with each other and our work is of significance in that they contribute to shed light on the basis mechanisms of collective phenomena beheld in the realistic social systems.

\section*{ACKNOWLEDGMENTS}\label{secfive}
C. Xu, J. Gao, H. R. Xiang, W. J. Jia, were supported partly by the National Natural Science Foundation of China (Grant No. 110107703). Z. G. Zheng was supported partly by the scientific Research Funds of Huaqiao University.

\section*{APPENDIX: THE OTT-ANTONSEN METHOD FOR THE RELAXATION DYNAMICS}
 \setcounter{equation}{0}
 \renewcommand\theequation{A\arabic{equation}}
 In this Appendix, we provide the Ott-Antonsen method that allows to describe the generalized frequency-weighted Kuramoto model in the low-dimensional invariant manifold, following this method, the density function $F(\theta, t, \omega)$ is sought in the form:
 \begin{equation}\label{equ:apa1}
 F(\theta, t, \omega)=\dfrac{g(\omega)}{2\pi}(1+F_{+}+F_{-}),
 \end{equation}
 where
 \begin{equation}\label{equ:apa2}
 F_{+}=\sum_{n=1}^{\infty}F_{n}(\omega, t)e^{in\theta},
 \end{equation}
 and $F_{-}=F_{+}^{*}$, with the additional ansatz
 \begin{equation}\label{equ:apa3}
 F_{n}=\alpha^{n}(\omega, t).
 \end{equation}
 Substituting the ansatz into the continuity equation, one obtain the equation for $\alpha(\omega, t)$
 \begin{equation}\label{equ:apa4}
 \dfrac{d\alpha}{d t}+i \omega \alpha+\dfrac{f(\omega)K}{2}(z\cdot \alpha^{2}-z^{*})=0,
 \end{equation}
with the order parameter,
\begin{equation}\label{equ:apa5}
 z(t)=\int g(\omega)\alpha^{*}(\omega, t)d\omega,
 \end{equation}
 as the similar analysis in the mean-field theory, we look for a steady state $z(t)=r e^{i\Omega t}$, for the constant order parameter $r$, and a group velocity $\Omega$. With a suitable change of the reference frame $\omega\rightarrow \omega+\Omega$ and putting $\dot{\alpha}=0$, we find $\alpha_{0}(\omega)$ a solution
 \begin{equation}\label{equ:apa6}
 \footnotesize
  \begin{cases}
\pm\sqrt{1-\left(\dfrac{\omega}{K r f(\omega+\Omega)}\right)^{2}}-\dfrac{i\omega}{K r f(\omega+\Omega)},\quad |\omega|\leq K r |f(\omega+\Omega)|,\\
  \\
  -\dfrac{i\omega}{K r f(\omega+\Omega)}\left[1-\sqrt{1-\left(\dfrac{K r f(\omega+\Omega)}{\omega}\right)^{2}}\right],\quad otherwise.
  \end{cases}
  \end{equation}
  Substituting Eq.~(\ref{equ:apa6}) into Eq.~(\ref{equ:apa5}) yields the self-consistency equations (Eq.~(\ref{equ:re16}) and Eq.~(\ref{equ:re17})) in the main text.

  To investigate the relaxation dynamics, we consider a weak perturbation from a stationary state,
  \begin{equation}\label{equ:apa7}\begin{aligned}
  z(t)=&r+\delta z(t),\\
  \alpha(t)=&\alpha_{0}(\omega)+\delta\alpha(\omega, t),
  \end{aligned}
  \end{equation}
  we obtain a linear equation for $\delta\alpha(\omega,t)$
 \begin{equation}\label{equ:apa8}
 \dfrac{d\delta\alpha}{dt}+i\omega\delta\alpha+\dfrac{K f(\omega)}{2}(2 r\alpha_{0}\delta\alpha+\alpha_{0}^{2}\delta z-\delta z^{*})=0,
 \end{equation}
 that must be solved self-consistently with
 \begin{equation}\label{equ:apa9}
 \delta z(t)=\int g(\omega)\delta \alpha^{*}(\omega,t)d\omega.
 \end{equation}
 Taking the Laplace transform of both sides of Eq.~(\ref{equ:apa8}), and reordering the terms then yields
 \begin{equation}\label{equ:apa10}
 \delta\alpha(s,\omega)=\dfrac{\delta \alpha(t=0)+\frac{K f(\omega)}{2}(\delta\alpha^{*}(s)-\alpha_{0}^{2}\delta z(s))}{s+i\omega+K f(\omega)r \alpha_{0}}.
 \end{equation}
 Substituting (\ref{equ:apa10}) into (\ref{equ:apa9}) leads to
  \begin{equation}\label{equ:apa11}
 \delta z(s)=\dfrac{B(s)}{1-KA(s)/2},
 \end{equation}
 where
 \begin{equation}\label{equ:apa12}
 B(s)=\int_{-\infty}^{\infty}g(\omega+\Omega)\dfrac{\delta\alpha(\omega, t=0)}{s+i\omega+K r f(\omega+\Omega)\alpha_{0}}d\omega
 \end{equation}
 and
 \begin{equation}\label{equ:apa13}
 A(s)=\int_{-\infty}^{\infty}g(\omega+\Omega)\dfrac{f(\omega+\Omega)(1-\alpha_{0}^{2})}{s+i\omega+K r f(\omega+\Omega)\alpha_{0}}d\omega,
 \end{equation}
 the order parameter $\delta z(t)$ is the inverse Laplace transform of $\delta z(s)$, It is obviously that when the system is the incoherent state, $\alpha_{0}\equiv 0$, and (\ref{equ:apa11}) is the same as Eq.~(\ref{equ:re41}) in the main text.


\begin{thebibliography}{99}
\bibitem{pikovsky2001}A. Pikovsky, M. Rosenblum and J. Kurths, {\em Synchronization: a Universal Concept in Nonlinear Sciences.} (Cambridge University Press, Cambridge, England, 2001).

\bibitem{strogatz2004} S. H. Strogatz, {\em Sync: How Order Emerges from Chaos in theUniverse, Nature and Daily Life} (Hyperion, New York, 2004).

\bibitem{arenas2008} A. Arenas, A. D{\'i}az-Guilera, J. Kurths, Y. Moreno and C. Zhou, Phys. Rep. {\bf 469}, 93 (2008).

\bibitem{breakspear2010} M. Breakspear, S. Heitmann and A. Daffertshofer, Front. Hum. Neurosci. {\bf 4}, 190 (2010).

\bibitem{Kuramoto1975} Y. Kuramoto, {\em in Lecture Notes in Physics}, Vol. 30, edited by H. Araki (Springer, New York, 1975).

\bibitem{acebron2005} J. A. Acebr{\'o}n, L. L. Bonilla, C. J. P{\'e}rez Vicente, F. Ritort and R. Spigler, Rev. Mod. Phys. {\bf 77}, 137 (2005).

\bibitem{rodrigues2016} F. A. Rodrigues, T. K. DM. Peron, P. Ji and J. Kurths, Phys. Rep. {\bf 610}, 1 (2016).

\bibitem{kuramoto1984} Y. Kuramoto, {\em Chemical Oscillations, Waves and Turbulence}. pp. 75-76 (Springer, Berlin, 1984).

\bibitem{zanette2005} D. H. Zanette, Europhys. Lett. (Europhysics Letters) {\bf 72}, 190 (2005).

\bibitem{paissan2007} G. H. Paissan and D. H. Zanette, Europhys. Lett. (Europhysics Letters), {\bf 77}, 20001 (2007). 

\bibitem{paissan2008} G. H. Paissan and D. H. Zanette, Phys. D {\bf 237}, 818--828 (2008).

\bibitem{vlasov2014} V. Vlasov, E. E. N. Macau and A. Pikovsky, Chaos {\bf 24}, 023120 (2014).

\bibitem{skardal2016}P. S. Skardal {\em et al.}, Phys. D {\bf 323}, 40--48 (2016). 

\bibitem{vlasov2015} V. Vlasov, A. Pikovsky and E. E. N. Macau, Chaos {\bf 25}, 123120 (2015).

\bibitem{yoon2015} S. Yoon {\em et al.}, Phys. Rev. E {\bf 91}, 032814 (2015).
\bibitem{coutinho2013} B. C. Coutinho {\em et al.}, Phys. Rev. E, {\bf 87}, 032106 (2013).

\bibitem{daido1996}H. Daido, Phys. D {\bf 91}, 24--66 (1996).

\bibitem{hong2011} H. Hong and S H. Strogatz, Phys. Rev. Lett. {\bf 106}, 054102 (2011). 

\bibitem{zhang2013} X. Zhang {\em et al.}, Phys. Rev. E {\bf 88}, 010802 (2013). 

\bibitem{leyva2013} I. Leyva {\em et al.}, Phys. Rev. E {\bf 88}, 042808 (2013). 

\bibitem{zhang2014}X. Zhang {\em et al.}, Sci. Rep. {\bf 4}, 5200 (2014). 

\bibitem{hu2014} X. Hu {\em et al.}, Sci. Rep. {\bf 4}, 7262 (2014).

\bibitem{zhou2015} W. Zhou {\em et al.}, Phys. Rev. E {\bf 92}, 012812 (2015). 

\bibitem{xu2016}C. Xu {\em et al.}, Sci. Rep. {\bf 6}, 21926 (2016). 

\bibitem{wang2011} H. Wang and X. Li, Phys. Rev. E {\bf 83}, 066214 (2011).

\bibitem{zhu2013} L. Zhu {\em et al.}, Phys. Lett. A {\bf 377}, 2749--2753 (2013). 

\bibitem{qiu2015} T. Qiu {\em et al.}, Sci. Rep. {\bf 5}, 18235 (2015). 

\bibitem{basnarkov2008}L. Basnarkov and V. Urumov, Phys. Rev. E {\bf 78}, 011113 (2008).

\bibitem{petkoski2013} S. Petkoski, D. Iatsenko, L. Basnarkov and A. Stefanovska, Phys. Rev. E {\bf 87}, 032908 (2013).

\bibitem{strogatz1991}S. H. Strogatz and R. E. Mirollo, J. Stat. Phys. {\bf 63}, 613--635 (1991). 

\bibitem{chiba2011} H. Chiba and I. Nishikawa, Chaos {\bf 21}, 043103 (2011).

\bibitem{chiba2015} H. Chiba, Ergodic Theory and Dynamical Systems, {\bf 35}, 762--834 (2015).

\bibitem{stanley1987}H. E. Stanley, {\em Introduction to Phase Transitions and Critical Phenomena, International Series of Monographs on Physics.} (Oxford University Press, New York, 1987).

\bibitem{strogatz1992}S. H. Strogatz, R. E. Mirollo and P. C. Matthews, Phys. Rev. Lett. {\bf 8}, 2730--2733 (1992). 

\bibitem{malmberg1964} J. H. Malmberg and C. B. Wharton, Phys. Rev. Lett. {\bf 13}, 184 (1964).

\bibitem{ott2008} E. Ott and T. M. Antonsen, Chaos {\bf 18}, 037113 (2008).

\bibitem{ott2009}E. Ott and T. M. Antonsen, Chaos {\bf 19}, 023117 (2009).

\bibitem{crawford1994} J. D. Crawford, J. Stat. Phys. {\bf 74}, 1047--1084 (1994).

\bibitem{crawford1999} J. D. Crawford and K. T. R. Davies, Phys. D {\bf 125}, 1--46 (1999).

\end{thebibliography}
\end{document}